\documentclass[twoside,slac_one]{revtex4}
\usepackage{graphicx}
\usepackage{fancyhdr}
\usepackage{amsmath} 
\usepackage{bm}
\usepackage{amsxtra}
\usepackage{amssymb}
\usepackage{amsthm}
\usepackage{latexsym}
\usepackage{lscape}

\begin{document}

\title{Measurement of the anomalous like-sign dimuon charge asymmetry with
9 fb$^{-1}$ of $p \bar{p}$ collisions}

%

\author{B. Hoeneisen}
\affiliation{Universidad San Francisco de Quito, Ecuador}

\begin{abstract}
The D\O\ Collaboration has performed a new 
measurement of the anomalous like-sign dimuon charge asymmetry with 
9 fb$^{-1}$ of $p \bar{p}$ collisions. In these proceedings I present
a short overview of the measurement that complements the slides presented
at the DPF-2011 Conference.
\end{abstract}

\maketitle

\thispagestyle{fancy}


\section{\label{proceedings} Introduction}

At the DPF-2011 Conference I presented a new
measurement, by the D\O\ Collaboration, of the
anomalous like-sign dimuon charge asymmetry 
of $p \bar{p}$ collisions with
9.0 fb$^{-1}$ of data. 
Since then the article has been accepted for
publication so I refer the reader
to \cite{PRD3} for full details of the 
measurement. Here I will only give a short
overview, and some personal comments. Complementary
to this note are the slides presented at the DPF-2011 Conference.

At the Fermilab Tevatron collider, $b$ quarks are produced mainly in $b \bar{b}$ pairs.
Therefore, to observe an event with two like-sign muons
from semi-leptonic $b$-hadron decay, one
of the hadrons must be a $B^0_d$ or $B^0_s$ meson that oscillates and decays to
a muon of charge opposite of that of the original $b$ quark.
The oscillation $B^0_q \leftrightarrow \bar{B}^0_q$ ($q = d$ or $s$) is described by
``box" Feynman diagrams that are sensitive to new particles
not directly accessible at the Tevatron.
The like-sign dimuon charge asymmetry from semi-leptonic decay of $b$-hadrons,
\begin{equation} 
A^b_\textrm{sl} \equiv \frac{N^{++}_b - N^{--}_b}{N^{++}_b + N^{--}_b},
\label{A}
\end{equation} 
has contributions from the semi-leptonic charge asymmetries
$a^d_\textrm{sl}$ and $a^s_\textrm{sl}$ of $B^0_d$ and $B^0_s$ 
mesons:
\begin{eqnarray}
A^b_\textrm{sl} & = & C_d a^d_\textrm{sl} + C_s a^d_\textrm{sl}. 
\label{Ab}
\end{eqnarray}

The D\O\ detector is well suited for this precision measurement
of $A^b_\textrm{sl}$
for the following reasons: (i) the initial $p \bar{p}$ state
is symmetric with respect to \textit{CP} conjugation; 
(ii) the solenoid and toroid magnetic fields are reversed
periodically, thereby canceling first order detector 
asymmetries; and (iii) the shielding between the central tracker
and the outer muon spectrometer is sufficient to reduce the
background from hadron punch-through to the 1\% level.

\section{\label{measurement} The measurement}
Two data sets are used for this measurement:
the \textit{inclusive muon set}, collected with single muon triggers,
has $n^+ + n^- = 2.04 \times 10^9$ muon candidates passing
strict quality selections; and the \textit{like-sign dimuon set},
collected with dimuon triggers, has
$N^{++} + N^{--} = 6.02 \times 10^6$ 
dimuons events with each muon passing the
same quality selections, and in addition the following dimuon
requirements: same charge sign, same associated vertex, and
a dimuon invariant mass greater than 2.8 GeV to suppress
events with the two muons coming from the same $b$-hadron decay
cascade. Counting inclusive muons, and like-sign dimuons, we
obtain the ``raw" asymmetries
\begin{eqnarray}
a & \equiv & \frac{n^+ - n^-}{n^+ + n^-} = (+0.688 \pm 0.002)\%, \label{a1} \\
A & \equiv & \frac{N^{++} - N^{--}}{N^{++} + N^{--}} = (+0.126 \pm 0.041)\%,
\end{eqnarray}
respectively.
These ``raw" asymmetries are corrected for kaon, pion and proton decay or
punch-through, and for the residual muon detector asymmetry after averaging
over the 4 solenoid-toroid polarity combinations. These corrections
are measured (as a function of the muon momentum transverse to the
$p \bar{p}$ beams, $p_T$) with the same data sets, by reconstructing
exclusive decays, with minimal use of simulation.
The main background asymmetry is due to kaon decay. Positive
kaons have a longer inelastic interaction length in the calorimeter
than negative kaons, and hence have more time to decay.
The resulting positive charge asymmetries from kaon decay are measured to
be $(+0.776 \pm 0.021)\%$ for $a$, and $(+0.633 \pm 0.031)\%$ for $A$.
The residual muon detector asymmetries are measured (reconstructing $J/\psi$'s
from central detector tracks) to be
$(-0.047 \pm 0.012)\%$ for $a$, and $(-0.212 \pm 0.030)\%$ for
$A$. Corrections due to pion decay and proton punch-through are
smaller. The charge asymmetries, corrected for background and detector
effects, are
\begin{eqnarray}
a - a_{\textrm{bkg}} & \equiv & (-0.034 \pm 0.042\textrm{ (stat)})\%, \label{a} \\
A - A_{\textrm{bkg}} & \equiv & (-0.276 \pm 0.067\textrm{ (stat)})\%. \label{b}
\end{eqnarray}

\begin{figure}[ht]
\centering
\includegraphics[width=80mm]{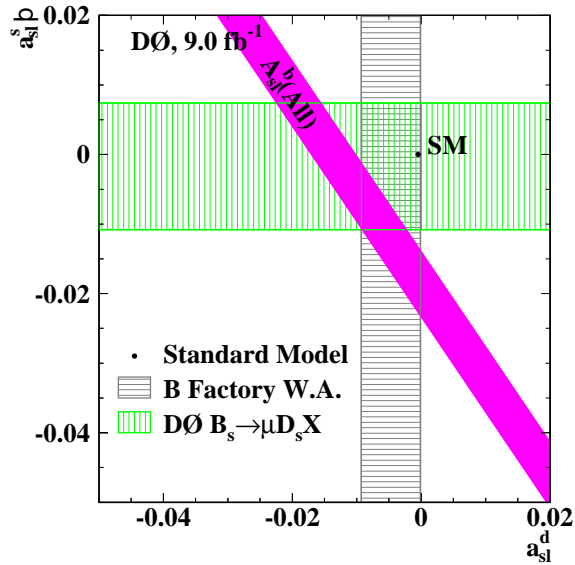}
\caption{Comparison of $A^b_\textrm{sl}$ in data with the
Standard Model prediction for $a^d_\textrm{sl}$ and $a^s_\textrm{sl}$.
Also shown are the measurements of $a^d_\textrm{sl}$~\cite{hfag} and
$a^s_\textrm{sl}$~\cite{asl-d0}. The bands represent the $\pm 1$
standard deviation uncertainties on
each individual measurement. } 
\label{1}
\end{figure}

We interpret these charge asymmetries as arising from \textit{CP}
violation in the mixing of $B^0_d$ and $B^0_s$ mesons.
To obtain $A^b_\textrm{sl}$, we divide the corrected asymmetries 
$a - a_{\textrm{bkg}}$ and $A - A_{\textrm{bkg}}$ by ``dilution
factors"
\begin{eqnarray}
c_b & = & +0.061 \pm 0.007, \\
C_b & = & +0.474 \pm 0.032,
\end{eqnarray}
respectively. These ``dilutions factors", obtained from 
simulation, are due to decays that are
not direct semi-leptonic $b \rightarrow \mu X$, e.g.
sequential decays $b \rightarrow c \rightarrow \mu X$,
decays with $b \rightarrow c \bar{c} q$ with $c \rightarrow \mu X$
or $\bar{c} \rightarrow \mu X$, decays of light mesons,
events with $c \bar{c}$, and events with $b \bar{b} c \bar{c}$.
The results are
\begin{eqnarray}
A^b_\textrm{sl} & = & (-1.04 \pm 1.30~({\rm stat}) \pm 2.31~({\rm syst})) \%, \\
A^b_\textrm{sl}	& = & (-0.808 \pm 0.202~({\rm stat}) \pm 0.222~({\rm syst})) \%,
\end{eqnarray}
respectively. The asymmetries $a$ and $A$ have correlated backgrounds.
Therefore, a more precise measurement of $A^b_\textrm{sl}$ can be obtained from
$A - \alpha a$. The parameter $\alpha = 0.89$ is chosen to minimize the total 
uncertainty of $A^b_\textrm{sl}$. The resulting final measurement is
\begin{eqnarray}
A^b_\textrm{sl}	& = & (-0.787 \pm 0.172~({\rm stat}) \pm 0.093~({\rm syst}))\%.
\label{Asl_final}
\end{eqnarray}
This result differs from the Standard Model prediction, 
\begin{equation}
A^b_\textrm{sl} = (-0.028^{+0.005}_{-0.006}) \%, 
\end{equation}
by 3.9 standard deviations. Equation (\ref{Ab}) with the result (\ref{Asl_final})
is show in Figure \ref{1}.

\begin{figure}[ht]
\centering
\includegraphics[width=0.50\textwidth]{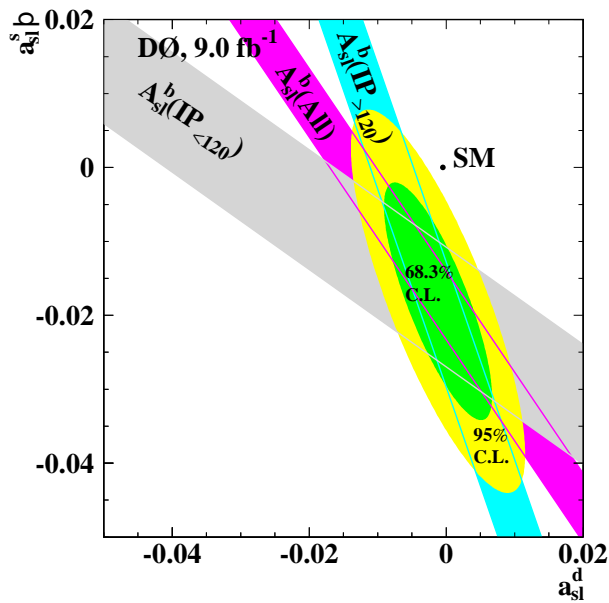}
\caption{Measurements of $A^b_\textrm{sl}$ with different muon impact parameter
selections in the $(a^d_\textrm{sl}, a^s_\textrm{sl})$ plane.
The bands represent the $\pm 1$ standard deviation uncertainties on each
individual measurement. The ellipses represents the 
$68.3\%$ and $95\%$ confidence-level contours of
$a^d_\textrm{sl}$ and $a^s_\textrm{sl}$ values obtained from the measurements
with IP selections. 
}
\label{2}
\end{figure}

To explore the origin of the charge asymmetry,
we perform measurements with muon impact parameter IP $> 120 \mu$m and
IP $< 120 \mu$m (for like-sign dimuons each muon is required to pass
the IP cut). IP is the distance of closest approach of the muon track
to the primary vertex projected onto
the plane transverse to the $p \bar{p}$ beams.
The coefficients $C_d$ and $C_s$ in (\ref{Ab})
depend on the IP cut, since for IP $> 120 \mu$m the $B^0_d$-meson
has a longer lifetime on average, and hence a greater probability
to oscillate. The results of these measurements are consistent with the
hypothesis of \textit{CP} violation in the mixing of $B^0_d$ and $B^0_s$
mesons with semi-leptonic decay asymmetries
\begin{eqnarray}
a^d_\textrm{sl} & = & (-0.12 \pm 0.51) \%,  \\
a^s_\textrm{sl} & = & (-1.81 \pm 1.04) \%.
\end{eqnarray}
These two asymmetries are correlated as shown in 
Figure \ref{2}.

\section{\label{comments} Comments}
The semi-leptonic charge asymmetry (\ref{Asl_final}), obtained
with 9.0 fb$^{-1}$ of data \cite{PRD3}, is in good agreement with
previous measurements by the D\O\ Collaboration with
1.0 fb$^{-1}$ \cite{PRD1} and 6.1 fb$^{-1}$ \cite{PRD2} of data.
These references \cite{PRD1, PRD2, PRD3} describe many cross-checks
of the measurement. Each one of these cross-checks could have
revealed inconsistencies, yet none have surfaced so far.

The ``raw" inclusive muon charge
asymmetry $a$ is dominated by background due to
the small value of the dilution factor $c_b$. Therefore (\ref{a})
serves as a ``closure test" of the measurements of the backgrounds
and detector asymmetries. Equation (\ref{a}) indicates that the
total uncertainty of all background and detector asymmetries (those that have
been explicitly considered, and even those that have not been imagined)
is less than approximately $\pm 0.042 \%$, which is
smaller than the statistical uncertainty of $A - A_{\textrm{bkg}}$.
The ``closure test" has been presented in \cite{PRD3} as a function
of transverse momentum $p_T$, and pseudo-rapidity $\eta$, and
good agreement is found.

Since background muons are mainly produced by decays of kaons 
and pions, their track parameters measured in the central
tracker and by the outer muon spectrometer can differ.
The background fractions therefore depend strongly on the
$\chi^2$ of the difference between these two measurements.
In test C of Table XV of \cite{PRD3} the $\chi^2$ cut is changed from
12 to 4 (for 4 degrees of freedom). The ``raw" charge asymmetry
$a$ ($A$) changes from +0.688\% to +0.325\% (+0.126\% to -0.361\%),
yet the measured value of $A^b_\textrm{sl}$ does not change
significantly (see \cite{PRD3} for details). Note that
$A$ changes sign with this reduction of background.

In Tables XV and XVI of \cite{PRD3} are presented 18
tests by varying the muon $p_T$, $\eta$ and $\phi$ ranges,
the muon quality selections, the triggers, the impact
parameter, the instantaneous luminosity, using only
one pair of solenoid-toroid magnet polarities, and different
data running periods. 
The $\chi^2$ of these $18+1$ measurements of
$A^b_\textrm{sl}$'s, taking account
of common events, is 17.1 for 18 degrees of freedom.
These tests indicate that the total uncertainty of
$A^b_\textrm{sl}$ is correct.

Applying the impact parameter cut IP $> 120 \mu$m 
reduces the kaon and pion decay backgrounds by 
factors 3 to 5, the ``raw" charge asymmetry  
$a$ ($A$) changes from +0.688\%	to -0.014\% (+0.126\% to -0.529\%),
yet the results are again consistent.
Note that both asymmetries $a$ and $A$ change sign with
this reduction of the backgrounds, and the measured
$A^b_\textrm{sl}$ from $a$ 
($A^b_\textrm{sl} = (-0.422 \pm 0.240 \textrm{ (stat)} \pm 0.121 \textrm{ (syst)}) \%$) 
and $A$ 
($A^b_\textrm{sl} = (-0.818 \pm 0.342 \textrm{ (stat)} \pm 0.067 \textrm{ (syst)}) \%$) 
are compatible.

It is a challenge to imagine a background or detector effect that
can make both corrected charge asymmetries (\ref{a}) and (\ref{b}) zero
simultaneously.
Finally, we note in Eq. (\ref{Asl_final}) that the uncertainty of
$A^b_\textrm{sl}$ is still dominated by statistics.

\section{\label{conclusion} Conclusions}
The D\O\ Collaboration has measured an anomalous
like-sign dimuon charge asymmetry that differs from
the standard model prediction by 3.9 standard deviations.

\bigskip 

\begin{thebibliography}{9}   

\bibitem{PRD3}
V.M.~Abazov \textit{et al.} (D0 Collaboration), arXiv:1106.6308 [hep-ex], 
accepted for publication in Phys. Rev. D.

\bibitem{hfag}
D.~Asner {\it et al.} (HFAG), arXiv:1010.1589 [hep-ex] (2010).

\bibitem{asl-d0}
V.M.~Abazov \textit{et al.} (D0 Collaboration), Phys. Rev. D {\bf 82}, 012003 (2010).

\bibitem{PRD1}
V.M.~Abazov, \textit{et al.} (D0 Collaboration), Phys.~Rev.~D~\textbf{74}, 092001 (2006).

\bibitem{PRD2}
V.M.~Abazov \textit{et al.} (D0 Collaboration), Phys. Rev. D~{\bf 82}, 032001 (2010).
V.M.~Abazov \textit{et al.} (D0 Collaboration),  Phys. Rev. Lett.~{\bf 105}, 081801 (2010).

\end{thebibliography}

\end{document}